\documentclass[11pt,a4paper]{article}

\usepackage{jcappub}

\usepackage{amsmath}
\usepackage{amssymb}
\usepackage{tensor}
\usepackage{latexsym}
\usepackage{enumerate}
\usepackage{bbm}
\usepackage{amsthm}
\usepackage{graphicx}
\usepackage{grffile}
\usepackage{caption}
\usepackage{subcaption}
\usepackage{float}
\usepackage{latexsym}
\usepackage{xcolor}

\usepackage[most]{tcolorbox}
\newtcolorbox{highlighted}{colback=yellow,coltext=red,breakable}
\DeclareMathAlphabet{\pazocal}{OMS}{zplm}{m}{n}

\newcommand{\ima}{i}

\newcommand{\be}{\begin{equation}}
\newcommand{\ee}{\end{equation}}
\newcommand{\bea}{\begin{eqnarray}}
\newcommand{\eea}{\end{eqnarray}}
\newcommand{\pd}{\partial}

\newcommand{\bc}{\begin{center}}
\newcommand{\ec}{\end{center}}
\newcommand{\abcd}{\alpha\beta\gamma\delta}
\newcommand{\ab}{\alpha\beta}

\newcommand{\cd}{\gamma\delta}
\newcommand{\fr}{\frac}

\title{Quantum cosmology of  Bianchi VIII, IX LRS geometries}
\author[a]{A. Karagiorgos}
\author[a]{T. Pailas}
\author[b]{N. Dimakis}
\author[a]{G.O. Papadopoulos}
\author[a]{Petros A. Terzis}
\author[a]{T. Christodoulakis}

\affiliation[a]{Department of Nuclear and Particle Physics,
Faculty of Physics, National and Kapodistrian University of Athens,
Athens 15784, Greece}
\affiliation[b]{Center for Theoretical Physics, College of Physical Science and Technology Sichuan University, Chengdu 6100064, China
}

\emailAdd{alexkarag@phys.uoa.gr}
\emailAdd{teopailas879@hotmail.com}
\emailAdd{nsdimakis@scu.edu.cn}
\emailAdd{gopapado@phys.uoa.gr}
\emailAdd{pterzis@phys.uoa.gr}
\emailAdd{tchris@phys.uoa.gr}

\abstract{In the present work we revisit the axisymmetric Bianchi VIII and IX models. At the classical level we reproduce the known analytic solution, in a novel way making use of two quadratic integrals of motion, the constraint equation, as well as a linear non-local integral of motion. These quantities correspond to two second rank Killing tensors and a homothetic vector field  existing on the relevant configuration space. On the corresponding phase space the two quadratic charges commute with the Hamiltonian constraint  but not among themselves. Thus, after turning these charges into operators we obtain two different solutions to the Wheeler DeWitt equation utilizing each of the quadratic operators. The homothetic vector is then used, as a natural guide line,  to define a normalizable conditional probability which assigns zero to the classically collapsed configurations. }


\arxivnumber{-}

\begin{document}
\maketitle

\flushbottom

\section{Introduction}

A first approach to Bianchi IX cosmological model has been initiated by Misner \cite{Misner}, Ryan \cite{ryan} and Belinskii, Khalatnikov and Lifshitz (BKL) \cite{bkl1,bkl2}. In these works use has been made of the well known Misner variables $(\alpha,\beta_+,\beta_-)$, where $\alpha$ is the scale factor measuring the overall volume of the spatial slice, while $\beta+$ and $\beta_-$ its anisotropy. The system can be interpreted  as a particle moving in a time-dependent potential well. Dynamical and numerical studies have been performed for its trajectory implying a possible chaotic behaviour \cite{Ryan0,berger,Berger2,Berger3,motter}. The fully anisotropic model created great debate in the literature regarding its integrability and whether - and in which sense - it exhibits chaos, we refer the interested reader to some basic works concerning these matters \cite{Integr0,Integr0a,Integr1,Integr2,Integr3,Integr4,Integr5,Integr6,Integr6a,Integr6b,Integr7,Integr8}. On the other hand, the Bianchi type VIII model has not attracted the same level of interest in the literature as  has the type IX with its closed hypersurfaces. However, there exist interesting studies and exact solutions that have been derived \cite{type8a,type8b,type8c}.

It is widely known that for specific classes of spatially homogeneous space-times \cite{RyanShep} the principle of symmetric criticality \cite{Palais,Torre} holds, i.e. a reduced action that results from the space-time symmetries can be used to generate correctly the equations of motion of the model, i.e. those obtained by imposing the symmetry requirement at the level of Einstein's Field Equations. The emanating Lagrangian possesses a finite number of degrees of freedom and it is the basis of the so called mini-superspace description. There have been a lot of studies making use of this property especially in what regards the symmetries that the reduced system possesses \cite{symc1,symc2,symc2a,tcjphysa,symc3,symc4,symc5,symc6,symc7,symc8}. This procedure is of special interest in what regards the quantization of the reduced system, since this can give a hindsight of what one could expect from the quantum version of the gravitational configuration. This gave rise to the field of quantum cosmology \cite{quant1,quant2,quant3,quant4,quant5,quant6,quant7,quant8,quant9,quant10}. Both Bianchi types VIII and IX admit such a description.

In the present work we have, at the classical level, re-acquired the solution to the Einstein's equations for the empty axisymmetric Bianchi Types VIII, IX model. The novelty is that, in order to find the solutions, we utilize the constants of motion inferred from the corresponding mini-supermetric, i.e. the conserved (local) charges produced by two Killing tensor fields (second rank) and a non-local corresponding to a homothetic Killing vector field (first rank), along with the quadratic constraint. A similar approach has been recently adopted  for Bianchi IX plus a cosmological constant \cite{ospage}. The difference here is that the empty classical case has one more second rank symmetry.

At the quantum level, we turn the quadratic charges into operators which commute with the quadratic constraint operator (defining the Wheeler-DeWitt equation) but not among themselves. We thus obtain, using each of the quantum charges as subsidiary condition on the wave function, two distinct quantum states (modulo integration constants). We show that, with the help of the homothetic vector, a conditional probability can be defined which assigns zero to the classically collapsed configuration.

The structure of this paper is as follows: In section \ref{section1}  we  present the minisuperspace of axisymmetric Bianchi VIII, IX models, study its symmetries and, through them, obtain the classical solutions. In section \ref{section2} we study the quantization of the model. Section \ref{section3} contains our analysis on the choice of an internal time variable with the help of the homothetic symmetry of the mini-superspace metric. Finally, in the last section we give our conclusions.

\section{Mini-superspace for axisymmetric Bianchi VIII and IX model} \label{section1}
For the general Bianchi models VIII and XI, a generic scale factor matrix can be diagonalized solely through kinematics and then the use of linear constraints dictates the vanishing of the shift vector (see e.g. \cite{Christodoulakis:2001mg,chrskarzam}). In the present work we focus in the case of axisymmetry; our starting point is, thus, the line element
\be\label{lineel}
ds^2=-N^2(t)dt^2+\gamma_{AB}\sigma^A_i\sigma_j^B dx^idx^j, \quad A,B,i,j=1,2,3
\ee
with  $\gamma_{AB}=diag(a^2(t),b^2(t),b^2(t))$ (note that the alternative choices turn out to be incompatible for VIII \cite{chrkor1994,chrskarzam}) and the one-forms corresponding to Bianchi VIII,  IX  symmetry group are given by
\bea
\sigma^1&=& dx  -k\sinh (k y) dz\\
 \sigma^2&=&\cos (x) dy-\sin (x) \cosh (k y) dz \\
\sigma^3&=& \sin(x)dy+\cos(x) cosh(k y)dz
\eea
with $k=1$ and $k=i$ for Bianchi  VIII and Bianchi IX respectively.

The Ricci scalar for this metric is
\bea
R&=&\frac{2 a''(t)}{a(t) N(t)^2}+\frac{4 a'(t) b'(t)}{a(t) b(t) N(t)^2}-\frac{2 a'(t) N'(t)}{a(t) N(t)^3}-\frac{a(t)^2}{2 b(t)^4}+\nonumber \\
& & \frac{4 b''(t)}{b(t) N(t)^2}-\frac{4 b'(t) N'(t)}{b(t) N(t)^3}+\frac{2 b'(t)^2}{b(t)^2 N(t)^2}-k^2\frac{2}{b(t)^2}
\eea
It is well known that the reduced Einstein Hilbert action provides the correct dynamics i.e. the inferred Lagrangian is valid. In order to enclose the entire dynamics in the kinetic part, and also simplify our calculations, we rescale the lapse as  $N(t)= \frac{2 b(t)^2 n(t)}{a(t)^3+k^2 4 a(t) b(t)^2}$ which effectively makes the potential independent of $a$ and $b$;  the final form of the Lagrangian is therefore  given by
 \be
\pazocal L=-\frac{2 a(t) \left( 4 k^2 b(t)^2+a(t)^2\right) a'(t) b'(t)}{b(t) n(t)}-\frac{a(t)^2 \left(4 k^2 b(t)^2+ a(t)^2\right) b'(t)^2}{b(t)^2 n(t)}-n(t).
 \ee
For the sake of simplicity we adopt, in what follows, the change to light-cone coordinates $w,u$
\be
(a,b)\mapsto (u,w): \quad a=\frac{w}{\sqrt u},\quad b=u.
\ee
Thus, the transformed Lagrangian and its corresponding mini-supermetric are written as
\be \label{Laguw}
L=-\frac{2 w(t)\left(4 k^2 u(t)^3+w(t)^2\right)}{n(t) u(t)^3} u'(t) w'(t)  -n(t)
\ee
\be\label{hypermetric}
G_{\ab}=
\left(
\begin{array}{cc}
 0 & -\frac{2 w^3}{u^3}-8 k^2 w \\
 -\frac{2 w^3}{u^3}-8 k^2 w & 0 \\
\end{array}
\right)
\ee
The above Lagrangian is singular and applying the Dirac-Bergmann \cite{dirac,Berg} algorithm for constrained systems we obtain the weakly vanishing Hamiltonian constraint
\be
\pazocal{H}=G^{\alpha\beta}p_\alpha p_\beta + 1 =
\frac{- u(t)^3}{2 w(t) \left(4 k^2 u(t)^3+w(t)^2\right)}p_u(t) p_w(t) +1\approx0
\ee
with $p_w(t)\equiv \frac{\partial L}{\partial w'(t)},\quad p_u(t)\equiv \frac{\partial L}{\partial u'(t)}$.
In order to reproduce the classical solution we can use possible extra symmetries of the above configuration space. The mini-supermetric $G_{\alpha \beta}$ has both first  and second order symmetries. Of the first kind there is only one  homothetic vector field and of the second there are  two second rank Killing tensor fields; overall we have
\paragraph{First-order:}
\be
\pazocal L_{\xi_h} G_{\alpha\beta}=G_{\alpha\beta}\Rightarrow\xi_h=\frac{u}{4}\pd_u+\frac{3w}{8}\pd_w
\ee
\paragraph{Second-order:}
\be\label{killfield}\nonumber
\nabla_\mu K_{
\nu \lambda }+\nabla_\lambda K_{\mu
\nu }+\nabla_
\nu K_{\lambda \mu}=0\Rightarrow
\ee
\be
K_1^{\mu\nu}= \left(
\begin{array}{cc}
 \frac{u^2}{4} & -\frac{u w \left(4 k^2 u^3-w^2\right)}{8 \left(4 k^2 u^3+w^2\right)} \\
 -\frac{u w \left(4 k^2 u^3-w^2\right)}{8 \left(4 k^2 u^3+w^2\right)} & \frac{w^2}{16} \\
\end{array}
\right)
K_2^{\mu\nu}=\left(
\begin{array}{cc}
 0 & \frac{u}{4 w \left(4 k^2 u^3+w^2\right)} \\
 \frac{u}{4 w \left(4 k^2 u^3+w^2\right)} & \frac{1}{4 w^2} \\
\end{array}
\right)
\ee
It has been shown in \cite{tcjphysa,scnonlo} that proper homothetic or conformal Killing vectors of the mini-superspace metric can be used to define non-local conserved charges that involve an explicit time dependence in the form of an integral of phase-space variables. In this particular case, the non-local conserved quantity generated by $\xi_h$ is
\be
\begin{split}
Q_h=\xi^{\alpha}(q) p_\alpha + \int\!\! n(t)  dt=& -\frac{w(t) \left(4 k^2 u(t)^3+w(t)^2\right) \left(3 w(t) u'(t)+2 u(t) w'(t)\right)}{4 n(t) u(t)^3} \\
& + \int \!\! n(t)d t .
\end{split}
\ee
It is easy to verify that the total derivative of the above expression vanishes on mass shell. At the same time, the two Killing tensors define of course the following local integrals of motion
\bea
Q_{K_1}=K_1^{\mu\nu}p_\mu p_\nu &=&\frac{w(t)^2 \left(4 k^2 u(t)^3+w(t)^2\right)}{4 n(t)^2 u(t)^6}\left[-16 k^2 u(t)^4 w(t) u'(t) w'(t)\right.\\
& + &  4 k^2 u(t)^3 w(t)^2 u'(t)^2+16 k^2 u(t)^5 w'(t)^2+4 u(t) w(t)^3 u'(t) w'(t)\nonumber\\
&+&\left. w(t)^4 u'(t)^2+4 u(t)^2 w(t)^2 w'(t)^2\right]\nonumber
\eea
\be
\begin{split}
Q_{K_2}=K_2^{\mu\nu}p_\mu p_\nu=\frac{u'(t) \left(4 k^2 u(t)^3+w(t)^2\right) }{n(t)^2 u(t)^6} \Big[& 4 k^2 u(t)^3 u'(t)\\
& +w(t)^2 u'(t)+2 u(t) w(t) w'(t)\Big].
\end{split}
\ee
The above conserved quantities are more than enough for the acquisition of the classical solution. What is more, the latter can be derived by purely algebraic means. Note that this is possible only because the non-local charge gives non trivial time dependence to the algebraic solution obtained when the four charges are solved for $(u, p_{u},w, p_w)$. Thus,  combining together $Q_h=0, \ Q_{K_1}=\kappa_1, \ Q_{K_2}=\kappa_2$ and $\pazocal{H}\approx0$ we arrive at the solution. For the non-local conserved charge it is enough to require $Q_h=0$ instead of $Q_h=$constant because it involves an indefinite integral in its expression. This system of equations result into two different branches for the solution depending on whether the value of $\kappa_2$ is  zero or non-zero.

\subsection{Case with \texorpdfstring{$\kappa_2\neq0$}{k2 not zero}}
The general system of equations under consideration is the following
\bea
n^2(t)-\frac{2 w(t) \left(4 k^2 u(t)^3+w(t)^2\right)u'(t) w'(t)}{u(t)^3}=0 & &\label{eqcon}\\
Q_{K_1}-\kappa_1=0 & &\label{eqqk1} \\
Q_{K_2}-\kappa_2=0 & & \label{eqqk2}\\
Q_h=0 & & \label{eqqhom} .
\eea
where  \eqref{eqcon} has been obtained from $\mathcal{H}\approx 0$ expressed in terms of velocities.
At first we solve the constraint for $n(t)$ and replace into the two  equations \eqref{eqqk1} and \eqref{eqqk2}. The result is the following set of equations
\begin{subequations}
\begin{equation}\label{eqq2}
\begin{split}
-16 k^2 u(t)^4 w(t)^2 u'(t) w'(t)+4 k^2 u(t)^3 w(t)^3 u'(t)^2 & \\
+16 k^2 u(t)^5 w(t) w'(t)^2- 8 \kappa_1 u(t)^3 u'(t) w'(t)+4 u(t) w(t)^4 u'(t) w'(t)+& \\
w(t)^5 u'(t)^2+4 u(t)^2 w(t)^3 w'(t)^2 &=0
\end{split}
\end{equation}
\begin{equation}  \label{eqq1}
4 k^2 u(t)^3 u'(t)+w(t)^2 u'(t)-2 \kappa_2 u(t)^3 w(t) w'(t)+2 u(t) w(t) w'(t)=0
\end{equation}
and of course the equation produced by the nonlocal integral of motion
\be\label{eqqh}
4 f(t) n(t) u(t)^3-w(t) \left(4 k^2 u(t)^3+w(t)^2\right) \left(3 w(t) u'(t)+2 u(t) w'(t)\right)=0
\ee
\end{subequations}
where with $f(t)$ is the term $\int n(t)d t$. We can now solve the equations (\ref{eqq1},\ref{eqqh}) algebraically with respect to the velocities $(u'(t),w'(t))$. If this solution is substituted into  eq. \eqref{eqq2}, a purely algebraic relation among $(u,w)$ is obtained:
\be
4 \kappa_1+16 k^4 u(t)^4-8 \kappa_2 k^2 u(t)^3 w(t)^2+ 16 k^2 u(t) w(t)^2+\kappa_2 u(t)^2 \left(\kappa_2 w(t)^4-4 \kappa_1\right) =0.
\ee
Furthermore, eq. \eqref{eqq1} being an 1-form in 2D  is necessarily closed, and thus can be written in the form $\omega(u,w)d\phi(u,w)=0$ defining one more
algebraic relation between $(u, w)$
\be
2 \sqrt{16 k^4-\kappa_2^2 \kappa_1}+\frac{\kappa_2 u(t) \left(\kappa_2 w(t)^2-4 k^2 u(t)\right)+8 k^2}{\sqrt{1-\kappa_2 u(t)^2}}=0
\ee
Since, we have not used any choice of time these two relations are not independent and define one relation between $(u(t),w(t))$, namely
\be
w(t)=\pm \frac{1}{\kappa_2}\left(\frac{ \pm 2 \sqrt{16-\kappa_1 \kappa_2^2} \sqrt{1-\kappa_2 u(t)^2}+4 \kappa_2 k^2 u(t)^2-8 k^2}{u(t)}\right)^{1/2}.
\ee
The lapse $n(t)$ is given from the constraint equation as
\begin{equation}
  \begin{split}
  n(t) =& \pm \frac{2}{\kappa_2 u(t)^3} \Big[ 2 \kappa_2 k^2 u(t)^2+4 k^2 \pm \left(\kappa_1 \kappa_2^2-16\right)\left[\left(\kappa_1 \kappa_2^2-16\right) \left(\kappa_2 u(t)^2-1\right)\right]^{-1/2} \\
  & \pm \left[\left(\kappa_1 \kappa_2^2-16\right) \left(\kappa_2 u(t)^2-1\right) \right]^{1/2}+ 2 k^2 \left(\kappa_2 u(t)^2-1\right) \left(\kappa_2 u(t)^2+2\right)\Big]^{1/2} u'(t) .
  \end{split}
\end{equation}
It can be seen that the last two expressions satisfy the Euler-Lagrange equations of Lagrangian \eqref{Laguw}. Thus, the final form of the solution (given in terms of the initial variables $(N(t), a(t), b(t)$) can be expressed in the time gauge $b(t)=t$ as
\paragraph{First branch.} We write
\begin{subequations}\label{solut1}
\begin{align}
a(t)^2&=-\frac{2 \left(k^2 \left(4-2 \kappa_2 t^2\right)+\sqrt{\kappa_2^2 \mu  \left(1- \kappa_2 t^2\right)}\right)}{\kappa_2^2 t^2},\\
N(t)^2&=
\frac{2 \kappa_2^2 t^4}{\left(\kappa_2 t^2-1\right) \left(2 k^2 \left(\kappa_2 t^2-2\right)-\sqrt{\kappa_2^2 \mu  \left(1-\kappa_2 t^2\right)}\right)},
\end{align}
\end{subequations}
where we have set $\mu=\frac{16}{\kappa_2^2}-\kappa_1$. In order to not have a signature change and keep $a(t)^2$ and $N(t)^2$ positive, the variable $t$ has to be restricted appropriately by the constants $\kappa_2$ and $\mu$. For example:
\begin{itemize}
  \item In the $k=1$ case one must require that
   \begin{equation}
     \kappa_2 t^2 > 2 \quad \text{and} \quad  - 4 \left(\kappa_2 t^2-2\right)^2< \kappa_2^2 \left(\kappa_2 t^2-1\right) \mu \leq 0.
   \end{equation}
   \item On the other hand, if $k=i$ then the following inequalities must hold
   \begin{equation}
     1< \kappa_2 t^2 <2 \quad \text{and} \quad - 4 \left(\kappa_2 t^2-2\right)^2< \kappa_2^2 \left(\kappa_2 t^2-1\right) \mu \leq 0 .
   \end{equation}
\end{itemize}
\paragraph{Second branch.} In this case we have the square root appearing with the opposite sign
\begin{subequations}\label{solut2}
\begin{align}
a(t)^2 &=-\frac{2 \left(k^2 \left(4-2 \kappa_2 t^2\right)-\sqrt{\kappa_2^2 \mu  \left(1-\kappa_2 t^2\right)}\right)}{\kappa_2^2 t^2},\\
N(t)^2&=
\frac{2 \kappa_2^2 t^4}{\left(\kappa_2 t^2-1\right) \left(2 k^2 \left(\kappa_2 t^2-2\right)-\sqrt{\kappa_2^2 \mu  \left(1-\kappa_2 t^2\right)}\right)}.
\end{align}
\end{subequations}
The requirement of positive $a(t)^2$ and $N(t)^2$ leads to the conditions:
\begin{itemize}
  \item If $k=1$, then you may have either
  \begin{equation}
    1< \kappa_2 t^2 <2 \quad \text{and} \quad - 4 \left(\kappa_2 t^2-2\right)^2> \kappa_2^2 \left(\kappa_2 t^2-1\right) \mu
  \end{equation}
  or
  \begin{equation}
    \kappa_2 t^2 > 2 \quad \text{and} \quad \mu \leq 0 .
  \end{equation}
  \item In the other case where $k=i$ we are led to
  \begin{equation}
    1< \kappa_2 t^2 <2 \quad \text{and} \quad \mu \leq 0
  \end{equation}
  or
  \begin{equation}
    \kappa_2 t^2 > 2 \quad \text{and} \quad - 4 \left(\kappa_2 t^2-2\right)^2> \kappa_2^2 \left(\kappa_2 t^2-1\right) \mu  .
  \end{equation}
\end{itemize}

The number of the constants appearing in the solutions is exactly what is expected due to the axisymmetry assumption: $2\times$(number of independent scale factors)-$2\times$(number of first class constraints)$\equiv2\times 2-2\times1=2$.
Both of them are  essential as it can be seen by the fact that the Wronskian matrix $W_{ij}=\frac{\partial S_i}{\partial x^j
}$, where $S_i$'s are the four dimensional scalar curvatures ($K=R_{\mu\nu\kappa\lambda}R^{\mu\nu\kappa\lambda}$, $\Box K$, $K_{,\mu}K_{,\nu}g^{\mu\nu}$) and $x^j=(t,\kappa_1,\kappa_2)$, has $\mathrm{rank}=3$; which means that $t,\kappa_1,\kappa_2$ can be expressed as functions of $S_i$' s.

\subsection{Special case with \texorpdfstring{$\kappa_2=0$}{k2 equal to zero}}
In this case the system of equations under consideration is simpler. Again we follow the same procedure by solving algebraically eqs. \eqref{eqq1} and \eqref{eqqh} with respect to the velocities $u'(t)$ and $w'(t)$, where in the former we have st of course $\kappa_2=0$. The substitution into \eqref{eqq2} yields the following simplified relation between $u$ and $w$
\begin{equation}
  w(t) = \pm \frac{\sqrt{-\kappa_1-4 u(t)^4}}{2 k \sqrt{u(t)}} .
\end{equation}
The  final solution can be neatly expressed in terms of $(a,b,N)$ without any particular choice of time:
\begin{subequations}\label{euclsol1}
\begin{align}
a(t)^2 &=-\frac{4 b(t)^4+\kappa_1}{4 k^2 b(t)^2},\\
N(t)^2 &=\frac{16 k^2 b(t)^4 }{4 b(t)^4+ \kappa_1}b'(t)^2,
\end{align}
\end{subequations}
which, when $4b(t)^4+\kappa_1>0$, we observe that it is a solution of Lorentzian signature if $k=i$. On the other hand, in the $k=1$ case we see that the signature is $(+,-,+,+)$, indicating a shift of roles between $t,x$ coordinates. When $4b(t)^4+\kappa_1<0$ the opposite situation occurs. The  constant $\kappa_1$ is essential for the corresponding geometries.

As a preparation for the quantum case we next investigate all the Abelian Poisson Bracket subalgebras formed by the quadratic charges and the Hamiltonian. Since the Poisson bracket of $Q_{K_1}$ and $Q_{K_2}$ is non-zero the only remaining possibilities are
\be\label{pbra}
\{ \pazocal{H},Q_{K_1}\}=0
\qquad \{ \pazocal{H},Q_{K_2}\}=0.
\ee
 We can thus hope for two different quantum states which will correspond to the quantum analogues of either $\pazocal{H}$ and $Q_{K_1}$ or $\pazocal{H}$ and $Q_{K_2}$.

Lastly, a word about the linear homothetic charge is pertinent: Its non-local character excludes any use of the corresponding quantum analogue in the entire configuration space. Of course, if one wanted to indulge  into the reduced phase space where the ``true" Hamiltonian governs the dynamics, then the nonlocal linear integral $Q_h$ would be an ideal candidate for the time variable to be factored out along with the conjugate momenta; thus arriving at  a Schr\"odinger-like wave equation for $\Psi$.

\section{Quantum solutions} \label{section2}
It is well known that the quantum analogue of the Hamiltonian constraint i.e. the Wheeler-DeWitt(WDW) equation does not in most cases suffice to determine the wave function up to an additive constant phase; yet this is an essential requirement in the context of quantum mechanics. Fortunately, in the presence of further symmetries, one can supplement the WDW equation with the eigenvalue equations constructed out of the quantum analogues of the symmetry generators. In the canonical analysis the first step is to assume the mapping
\begin{equation*}
  \quad \{ \ , \  \}\rightarrow-i  [  \ , \  ]
\end{equation*}
between Poisson brackets and the commutators. A way for this o be realized is by assigning differential operators to momenta
\begin{equation*}
  p_n \mapsto \widehat{p}_n = -\ima\, \frac{\partial}{\partial n}, \quad p_\alpha \mapsto \widehat{p}_\alpha = - \ima \, \frac{\partial}{\partial q^\alpha},
\end{equation*}
while the positions are considered to act multiplicatively. In order to resolve the factor ordering problem of the Kinetic term of $\mathcal{H}$, we choose the conformal Laplacian (or Yamabe operator),
\begin{equation}\label{Hamoperator}
  \widehat{\mathcal{H}} = -\frac{1}{2 \mu} \partial_\alpha \left(\mu G^{\alpha\beta}\partial_\beta \right)+ \frac{d-2}{8(d-1)} \mathcal{R} + 1,
\end{equation}
where $\mu(q)= \sqrt{|\det{G_{\alpha\beta}}|}$, $\partial_\alpha = \frac{\partial}{\partial q^\alpha}$, $\mathcal{R}$ is the Ricci scalar and $d$ the dimension of the mini-superspace. Moreover, classical symmetries produced by linear vector and Killing tensor fields are naturally transcribed
 at the quantum level by  assigning to $Q_I$ the general expression for linear first order, Hermitian operators \cite{Dim1} and to $K_J$ a psedo-Laplacian operator \cite{benenti2002}; thus the corresponding forms are respectively
\begin{equation}\label{firstordop}
  \widehat{Q}_I = -\frac{\ima}{2\mu} \left( \mu \xi_I^{\alpha} \partial_\alpha + \partial_\alpha (\mu \xi_I^{\alpha})\right) = -\ima\, \xi_I^\alpha \partial_\alpha
\end{equation}
\be \label{Kop}
\widehat{K}_J=-\frac{1}{\mu}\partial_\alpha\left[\mu K_J^{\alpha\beta}\partial_\beta\right].
\ee
In our case the linear homothetic symmetry produces a non-local charge and thus it is not immediately usable in the four-dimensional phase space , spanned by $(a,b,p_a,p_b)$. Therefore, we turn to the quadratic charges produced by the two Killing tensors: one can easily verify that the relevant commutators are zero
\be\label{comm}
[\widehat{\mathcal H}, \widehat K_1]\Psi(a,b)=0\qquad [\widehat{\mathcal H}, \widehat K_2]\Psi(a,b)=0.
\ee
Note that the above relations hold for any $\Psi(a,b)$ irrespectively of whether it solves the Hamiltonian constraint or not. We thus distinguish the two Abelian quantum sub-algebras $(\widehat{\mathcal H}, \widehat K_1)$ and  $(\widehat{\mathcal H}, \widehat K_2)$.

\subsubsection*{Sub-algebra $(\widehat{\mathcal H}, \widehat K_1)$}
These operators give the following set of eigenvalue equations for the wave function
\be
 \pazocal {\widehat H}\Psi(u,w)=0\Rightarrow\nonumber
 \ee
 \be\label{hamcon}
\frac{u^3}{2 w \left(4 k^2 u^3+w^2\right)}\partial_u\partial_w\Psi(u,w)+\Psi(u,w)=0,
\ee

\be
\widehat K_1\Psi(u,w)=\kappa_1\Psi(u,w)\Rightarrow\nonumber
\ee
 \bea\label{kappaone}
16 \kappa_1\Psi(u,w)+w\partial_w\Psi(u,w)+w^2\partial_w\partial_w\Psi(u,w)+4 u\partial_u\Psi(u,w)+& &\nonumber\\
-4uw\frac{4k^2u^3-w^2}{4k^2u^3+w^2}\partial_u\partial_w\Psi(u,w)+4u^2\partial_u\partial_u\Psi(u,w)=0.& &
\eea
where $\kappa_1$ is the constant appearing in the classical solution. As a first step in the solution procedure, we solve the Hamiltonian constraint with respect to $\partial_w\partial_u\Psi(u,w)$, and replace  into \eqref{kappaone}.
In order to make the final  equations simpler and create a separable set of equations we use the  transformation
\be
\Psi(u,w)=f\left(\frac{u}{w^2}\right) g\left(u w^2\right),\qquad u= \sqrt{x} \sqrt{y},\qquad w= \frac{\sqrt[4]{y}}{\sqrt[4]{x}},
\ee
thus ending up with the following set of equations
\be \label{interm1}
\frac{y^2 g''(y)}{4 g(y)}+\frac{y g'(y)}{4 g(y)}+k^2 y=- \lambda^2
\ee
\be \label{interm2}
\frac{x^2 f''(x)}{2 f(x)}+\frac{x f'(x)}{2 f(x)}+\kappa_1 -\frac{1}{2 x^2}=\lambda^2.
\ee
The general solution to the above system is
\be\label{eigen2}
\Psi (u,w)=f\left(\frac{u}{w^2}\right) g\left(u w^2\right)
\ee
with
\be
f(\frac{u}{w^2})=c_1\Gamma \left(\sqrt{-\kappa_1}+1\right) I_{\sqrt{-\kappa_1}}\left(\frac{w^2}{u}\right)+c_2 \Gamma \left(1-\sqrt{-\kappa_1}\right) I_{-\sqrt{-\kappa_1}}\left(\frac{w^2}{u}\right),
\ee
\be
g(u w^2)=c_3 \Gamma \left(1-2 i \sqrt{\kappa_1}\right) J_{-2 i \sqrt{\kappa_1}}\left(4 k w\sqrt{u}\right)+c_4 \Gamma \left(1+2 i \sqrt{\kappa_1}\right) J_{2 i \sqrt{\kappa_1}}\left(4 k w\sqrt{u}\right),
\ee
where $c_1, \ c_2, \ c_3$ and $c_4$ are integration constants. The coupling constant $\lambda$ appearing in \eqref{interm1} and \eqref{interm1} has been fixed by the quadratic constraint relation at the value $\lambda = \pm \left(\frac{\kappa_1}{2}\right)^{1/2}$.

\subsubsection*{Sub-algebra $(\widehat{\mathcal H}, \widehat K_2)$}
The corresponding equations emerge from the quantum  Hamiltonian constraint given by \eqref{hamcon} and the action of $\widehat K_2$ on the wave function
\be
\widehat K_2\Psi(a,b)=\kappa_2\Psi(a,b)\Rightarrow\nonumber
\ee
\be\label{ktwo}-\frac{u \pd_u\pd_w\Psi(u,w)}{2 w \left(4 k^2 u^3+w^2\right)}-\frac{4 \kappa_2 w^3 \Psi(u,w)+w \pd_w\pd_w\Psi(u,w)-\pd_w\Psi (u,w)}{4 w^3}=0
\ee
Once more we solve \eqref{hamcon} with respect to  $\partial_w\partial_u\Psi(u,w)$, and substitute into the above equation, which is then reduced to
\be
\left(\frac{1}{u^2}-\kappa_2\right) \Psi(u,w)+\frac{\pd_w\Psi (u,w)-w  \pd_w\pd_w\Psi(u,w)}{4w^3}=0,
\ee
with the following general solution:
\be
\Psi(u,w)=d_2(u) \sin \left(\frac{w^2 \sqrt{ \kappa_2 u^2-1}}{u}\right)+d_1(u) \cos \left(\frac{w^2 \sqrt{ \kappa_2 u^2-1}}{u}\right).
\ee
The form of $d_2(u)$ and $d_1(u)$ is determined  by substituting the above expression for $\Psi(u,w)$ into equation \eqref{hamcon} and, in the resulting equation, setting to zero the coefficients of the above sin and cos . The final result is given by
\begin{equation}\label{sol2uw}
  \Psi(u,w)= \frac{u}{\sqrt{\kappa_2 u^2-1}} \left( c_5 e^{\frac{i \sqrt{\kappa_2 u^2-1} \left(4 k^2 u+\kappa_2 w^2\right)}{\kappa_2 u}} + c_6 e^{-\frac{i \sqrt{\kappa_2 u^2-1} \left(4 k^2 u+\kappa_2 w^2\right)}{\kappa_2 u}} \right)
\end{equation}
where $c_5$, $c_6$ are integration constants.

\section{Interpretation through Homothetic time} \label{section3}

Once we have secured the existence of quantum states defined  up to constants, we now must turn to the question of their interpretation. This is a well known, difficult problem in quantum cosmology (for a comprehensive review of the various approaches    see e.g. \cite{isham1994,Kuchar}). The main problem is the reparametrization invariance of the classical theory which is transcribed at the quantum level through the Wheeler DeWitt equation.

In the usual quantum mechanics normalizability of the states is expected by integration over  the configuration space variables but  not over time. The problem which arises in quantum cosmology is that any combination of the configuration space variables, say $\phi(a,b)$	in our case, must be allowed to be considered as time; consequently, in the definition of probability we must take into account this fact. In this respect two main approaches have been developed.

One consists in selecting the combination which is to represent time before quantization, thus arriving at a reduced phase space where the dynamics is governed by the "true" Hamiltonian through a Schr\"odinger like equation (see e.g. \cite{Kunstater,Barvi}). In the present case this would ideally fit with the use of the existing homothetic vector and it's conjugate to define time and energy.

The second scheme involves a selection of time after the quantization has been performed and the subsequent definition of a conditional probability on the complementary configuration space. In both approaches a really serious problem is the justification of the particular choice of time employed. It is natural to seek this justification in existing classical structures   of the configuration manifold.

In what follows we adopt the second point of view.  Since the two quadratic charges have already been used in the previous section for the derivation of our solutions, it is reasonable to turn to the linear homothetic charge for a justification of the choice for time.

Given that the dimension of the configuration space is two,  the corresponding to $\xi_h^{\mu}$ one form $\xi_{h\mu}\equiv G_{\mu\nu}\xi_h^\nu$ is necessarily closed, and  we thus expect that it can be brought into the form
\be
\xi_{h \mu}=-\omega(u,w) d f(u,w).
\ee
Indeed, it is straightforward to verify that the correct expressions are:
\be
\omega(a,b)=\frac{3 \left(4 k^2 u^3 w^2+w^4\right)}{4 u^3 f^{(1,0)}(u,w)} \qquad f(u,w)=f(u^{3/2} w).
\ee
Our proposal is now to select as natural time variable the quantity $\tau=u^{3/2} w$. Thus, the orthonormal coordinate system is composed by the following two variables
\be
(\tau, \chi)=(u^{3/2} w,\frac{\sqrt{u}}{\sqrt[3]{w}}), \quad (u,w)=\left(\sqrt[3]{\tau } \chi ,\sqrt{\fr{ \tau }{ \chi^3}}\right),
\ee
in which the transformed minisuperspace metric is diagonal
\be
G_{\mu\nu}(\tau,\chi)=
\left(
\begin{array}{cc}
 -\frac{2 \left(4 k^2 \chi ^6+1\right)}{3 \tau ^{2/3} \chi ^8} & 0 \\
 0 & \frac{6 \tau ^{4/3} \left(4 k^2 \chi ^6+1\right)}{\chi ^{10}} \\
\end{array}
\right)
\ee
with a corresponding Ricci scalar $\mathcal{R}=-\frac{24 k^2 \chi ^{14}}{\tau ^{4/3} \left(4 k^2 \chi ^6+1\right)^3}$. We see that there are curvature singularities of the configuration manifold described by $(\tau=0,\, \text{any}\, \chi)$  in the $VIII$ model and by $(\tau=0,\, \text{or}\, \chi=(\frac{1}{2})^{1/3})$ for the $IX$ model. If we notice that, in the initial variables $a,\,b$, $\tau=a b^2,\,\chi=(\frac{b}{a})^{1/3}$ and take into account the classical solutions, we conclude that $\tau=0$ corresponds to the 4-dim space-time curvature singularity, while $\chi=(\frac{1}{2})^{1/3}$ is never attained on the classical orbits. This singularity is brought in our configuration space geometry by the rescaling of the initial lapse $N(t)$ and thus corresponds to the vanishing of the Ricci scalar of the three dimensional spatial slice; thus never occurring for the considered models. Therefore, the range of integration for the variable $\chi$ will be initially taken $(\frac{1}{2})^{1/3}\,<\,\chi\,<\,\infty$ when defining the conditional probability.

The probability density is defined using the natural measure
\be
\mu(\tau,\chi)\equiv \sqrt{|\det{G_{\mu\nu}(\tau,\chi)}|}=\frac{2 \sqrt[3]{\tau } \left(4 k^2 \chi ^6+1\right)}{\chi ^9}
\ee
as
\be
\rho(\tau,\chi)\equiv \mu(\tau,\chi)\Psi(\tau,\chi)\Psi^*(\tau,\chi)
\ee

If the integral over $\chi$ of the above function converges, say to some $\rho_0(\tau)$, then we can define the conditional probability of the quantum state of  the universe to be in the configuration interval $(\tau,\chi) \rightarrow (\tau,\chi+\text{d} \chi)$ as
\be
P_\tau(\chi) \equiv \frac{\rho(\tau,\chi)}{\rho_0(\tau)}
\ee

Let us now apply this reasoning to the case of the second algebra. There are two separate regions depending on the value of $\kappa_2 u^2-1= \kappa_2 \tau ^{2/3} \chi ^2-1$ into the exponential function being positive or negative. The ensuing state is normalizable in the above conditional probability density; since $\int _{(\frac{1}{2})^{1/3}}^\infty P_\tau(\chi) \ \textrm{d} \chi=1$.  	

\subsection{Second solution}

The second algebra yields solution \eqref{sol2uw}, which in the $\chi, \tau$ variables reads
\be \label{sol2}
\begin{split}
\Psi(\tau,\chi)=\frac{\chi \sqrt[3]{\tau }   }{\sqrt{ \kappa_2 \tau ^{2/3} \chi ^2-1}} \Bigg[ & c_5\textrm{e}^{ \left(\frac{i \sqrt{\kappa_2 \tau ^{2/3} \chi ^2-1} \left(\kappa_2 \tau ^{2/3}+4 k^2 \chi ^4\right)}{\kappa_2 \chi ^4}\right)} \\
&+c_6\textrm{e}^{ \left(-\frac{i \sqrt{\kappa_2 \tau ^{2/3} \chi ^2-1} \left(\kappa_2 \tau ^{2/3}+4 k^2 \chi ^4\right)}{\kappa_2 \chi ^4}\right)}\Bigg],
\end{split}
\ee
\paragraph{In the first case} we consider that $\kappa_2 \tau ^{2/3} \chi ^2<1$ and the solution has a decay form. The wave function is in this case
\be
\psi_{dec}(\tau,\chi)=\frac{ \sqrt[3]{\tau } \chi  \exp \left(-\frac{\sqrt{1-\kappa_2 \tau ^{2/3} \chi ^2} \left(4 k^2 \chi ^4+\kappa_2 \tau ^{2/3}\right)}{\kappa_2 \chi ^4}\right)}{\sqrt{1-\kappa_2 \tau ^{2/3} \chi ^2}}
\ee
where, for the sake of normalizability, we have kept only the term which vanishes when $\chi$ tends to zero. Therefore, the probability density is
\be
\rho(\tau,\chi)=\frac{2  \tau  \left(4 k^2 \chi ^6+1\right) \exp \left(-\frac{2 \sqrt{1-\kappa_2 \tau ^{2/3} \chi ^2} \left(4 k^2 \chi ^4+\kappa_2 \tau ^{2/3}\right)}{\kappa_2 \chi ^4}\right)}{\chi ^7 \left(1- \kappa_2 \tau ^{2/3} \chi ^2\right)}
\ee
which tends to zero for  $\tau=0$ and for $\chi=2^{-1/3}$ for Bianchi type IX.

\begin{figure}[ht!]
    \centering
    \begin{subfigure}[b]{0.45\textwidth}
        \centering
        \includegraphics[trim=0.1cm 0 0.1cm 0, clip,width=1\textwidth]{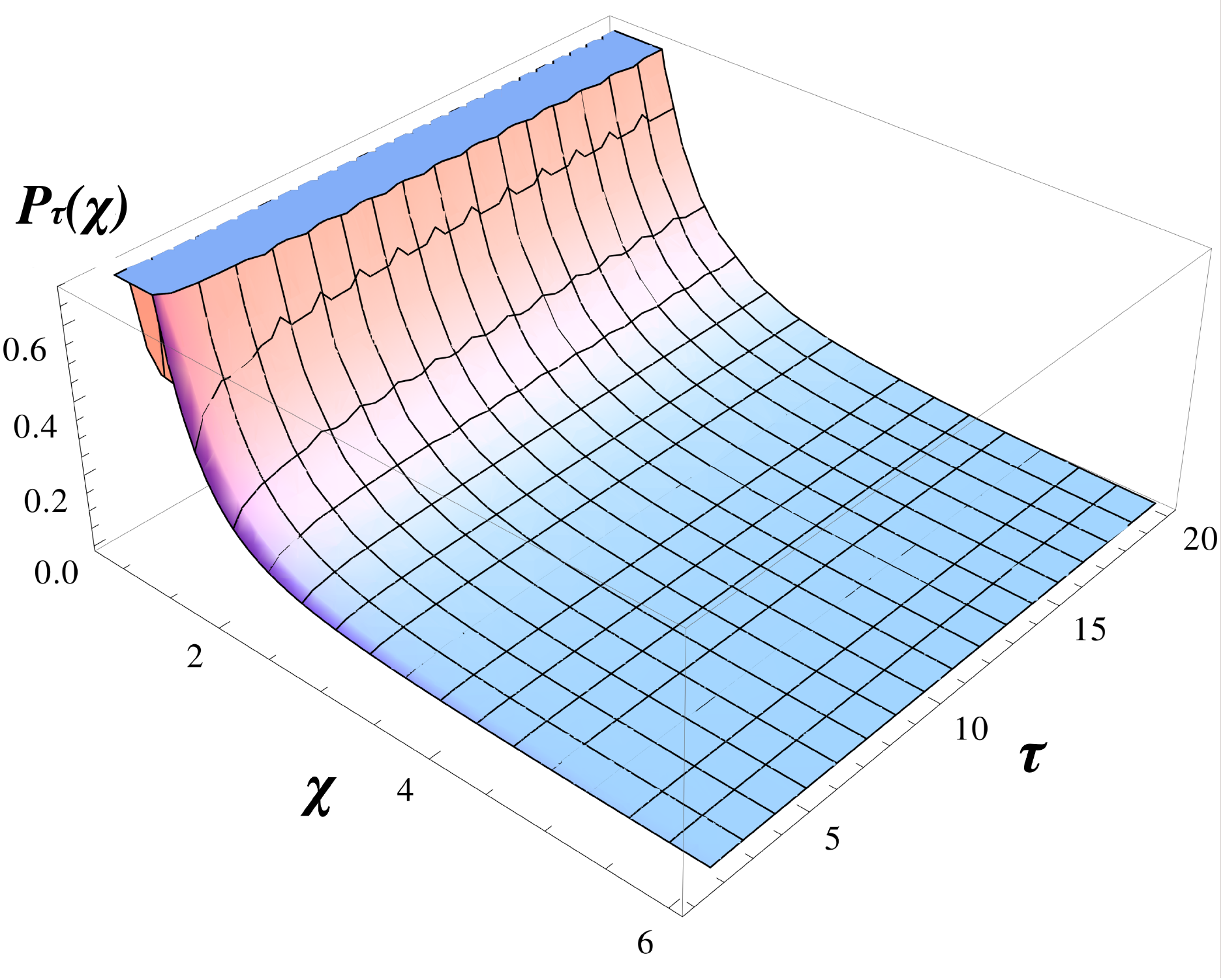}
        \caption{Bianchi VIII}
    \end{subfigure}%
    ~
    \begin{subfigure}[b]{0.45\textwidth}
        \centering
        \includegraphics[trim=0.1cm 0 0.1cm 0, clip,width=1\textwidth]{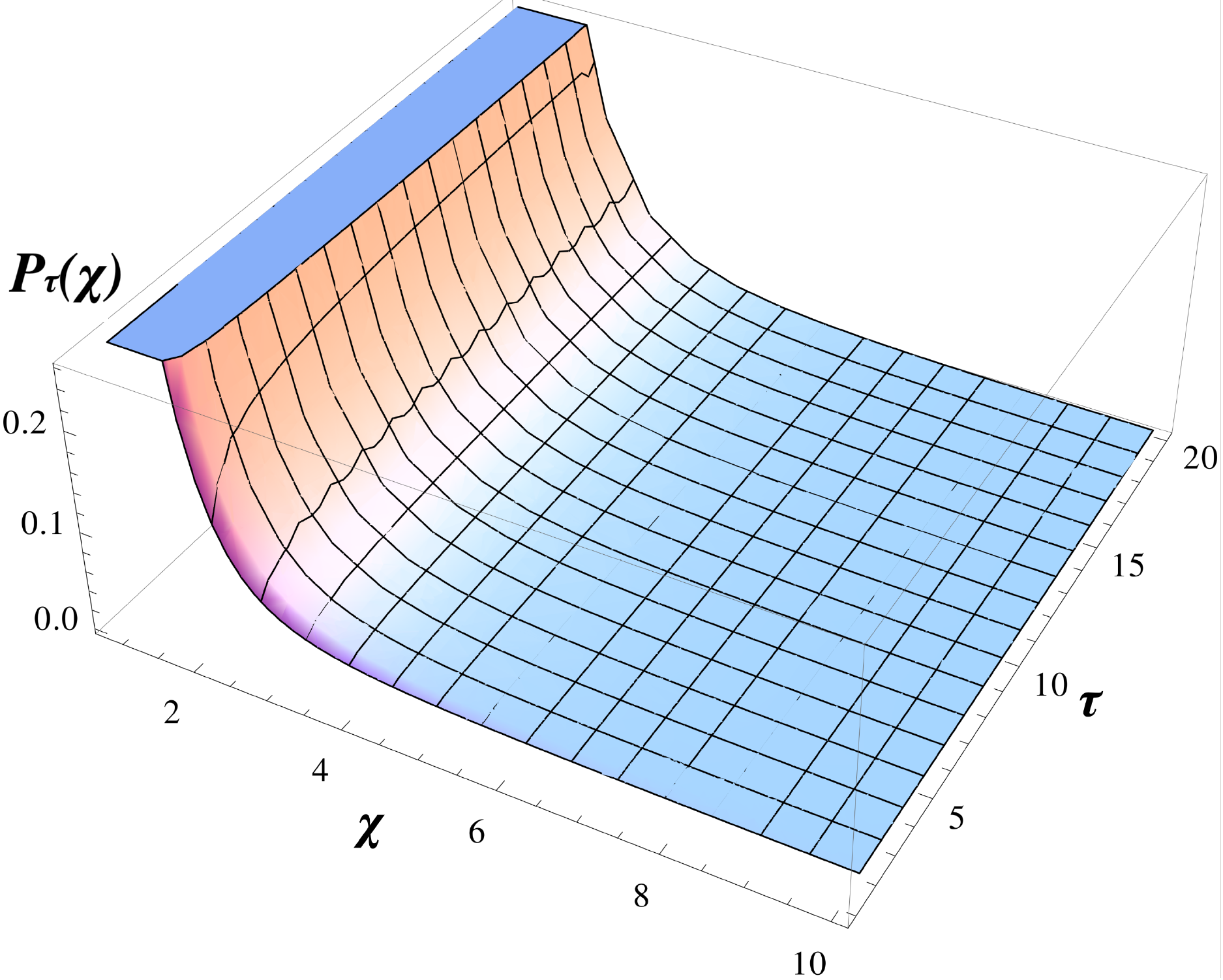}
        \caption{Bianchi IX}
    \end{subfigure}
    \caption{In these diagrams we plot the normalized probability density  in terms of $\chi$ and  $\tau$ for $\kappa_2=1$. }\label{probden}
\end{figure}

\paragraph{In the second case}
  $\kappa_2 \tau ^{2/3} \chi ^2>1$; now both terms oscillate. We consider  the positive frequency mode since we want only outgoing waves. Thus, by setting  $c_5=1$ and $c_4=0$. the wave function becomes
\be
\psi_1(\tau,\chi)=\frac{\chi \sqrt[3]{\tau }   \exp \left(-\frac{i \sqrt{\kappa_2 \tau ^{2/3} \chi ^2-1} \left(\kappa_2 \tau ^{2/3}+4 k^2 \chi ^4\right)}{\kappa_2 \chi ^4}\right)}{\sqrt{ \kappa_2 \tau ^{2/3} \chi ^2-1}}
\ee
The above defined $\rho_0(\tau)$ is given by
\be
\rho_0(\tau)=\frac{1}{3} \tau  \left(\left(4 k^2+\kappa_2^3 \tau ^2\right) \ln \left(\frac{ \kappa_2^3 \tau ^2}{\left(\kappa_2 \tau ^{2/3}-2^{2/3}\right)^3}\right)-3\sqrt[3]{2}\kappa_2\tau ^{2/3} (\sqrt[3]{2} \kappa_2\tau ^{2/3}+1)-4\right)
\ee
whose existence and finite behaviour is guaranteed by the inequality for this case.
The corresponding conditional probability density is
\be
P_\tau(\chi)\equiv\frac{\rho(\tau,\chi)}{\rho_0(\tau)}=\frac{2 \tau  \left(4 k^2 \chi ^6+1\right)}{\rho_0(\tau)(2 \kappa_2 \tau ^{2/3} \chi ^9-\chi ^7)}
\ee
which vanishes for the classically collapsed configurations designated by $\tau=0$.
In figure \eqref{probden} we plot the normalized probability density $P_\tau(\chi)$ with respect to $\tau$ and $\chi$. Note that the depicted  ranges for $\tau$ and $\chi$ are those allowed by the basic inequality above assumed.

\subsection{First solution}
In this case the wavefunction is given by  \eqref{eigen2} which
expressed in $\tau, \chi$ variables reads
\begin{align}
  f_1 (\tau,\chi) & = C_1 I_{-\sqrt{-\kappa_1}}\left(\frac{\tau^{2/3}}{\chi^4}\right) + C_2 I_{\sqrt{-\kappa_1}}\left(\frac{\tau^{2/3}}{\chi^4}\right) \\
  f_2 (\tau,\chi) & = C_3 J_{-2 \sqrt{-\kappa_1}} \left( 4 k \frac{\tau^{2/3}}{\chi}\right) + C_3 J_{2 \sqrt{-\kappa_1}} \left( 4 k \frac{\tau^{2/3}}{\chi}\right)
\end{align}
where we have absorbed the Gamma functions into  the constants $C_i$.The situation is considerably more involved since products of the Bessel functions appear in every term of $\Psi^*\Psi$. We only give the limitations on $\kappa_2$ imposed by the demand of normalizablity: The only dangerous   point is at $\chi\rightarrow +\infty$ with fixed $\tau$, where the arguments become small and tend to zero. Both $J_\mu$ and $I_\mu$ are problematic at zero if $\mu$ is negative while at the same time not an integer, thus $\mu \notin \mathbb{R}_{-}-\mathbb{Z}_{-}$. For the other values $J_\mu(0)=I_{\mu}(0)=0$, unless $\mu=0$ for which $J_0(0)=I_0(0)=1$.
\begin{enumerate}
  \item If $\kappa_1<0$ then $\mu$ is real and distinct approaches can be followed:
    \begin{enumerate}
      \item \label{case1a} Eliminate the branches of the solution that have a negative index by setting $C_1 = C_3=0$ and have a wave function
          \begin{equation}
            \Psi (\tau,\chi) = C_0 I_{\sqrt{-\kappa_1}}\left(\frac{\tau^{2/3}}{\chi^4}\right) J_{2 \sqrt{-\kappa_1}} \left( 4 k \frac{\tau^{2/3}}{\chi}\right), \quad \kappa_1 <0.
          \end{equation}
      \item Keep both branches, but restrict $\kappa_1$ so that $\sqrt{-\kappa_1} \in \mathbb{Z}_+$. However, in this case, $J_\mu$ and $J_{-\mu}$ seize to be independent since $J_{-\mu} =(-1)^\mu J_\mu$ when $\mu$ is an integer. The same is also true for the $I_\mu$ and $I_{-\mu}$ when $\mu \in \mathbb{Z}$, $I_{-\mu}=(-1)^\mu I_\mu$. Thus, we end up again with the same wave function as in case \ref{case1a}, but with a different condition on $\kappa_1$.
          \begin{equation}
            \Psi (\tau,\chi) = C_0 I_{\sqrt{-\kappa_1}}\left(\frac{\tau^{2/3}}{\chi^4}\right) J_{2 \sqrt{-\kappa_1}} \left( 4 k \frac{\tau^{2/3}}{\chi}\right), \quad \sqrt{-\kappa_1} \in \mathbb{Z}_+.
          \end{equation}
    \end{enumerate}

  \item If $\kappa_1>0$ then the indexes are imaginary, $\mu = \ima \alpha$, $\alpha \in \mathbb{R}$. The resulting Bessel functions are bounded, but the limit at zero cannot be calculated. It is a limit of the form
      \begin{equation}
        0^{\ima \alpha} =e^{\ima \alpha \ln(0)} = \cos(\alpha \ln(0))+ \ima \sin(\alpha \ln(0))= \cos (\infty) + \ima \sin(\infty),
      \end{equation}
      it fluctuates on the complex plane. However, whatever one might choose here the wave function is going to be bounded as $\chi\rightarrow +\infty$.
\end{enumerate}
Thus, given the above described allowed values for the indexes the $\Psi^*\Psi$ would be finite and a similar qualitative analisis to the previous case will hold.

\section{Discussion}
In the present work we have studied the mini-superspace models of  Bianchi type VIII and IX LRS geometries.

 At the classical level we have reproduced the solution space through the use of the symmetries of the configuration space manifold. Specifically, these symmetries consist of a homothetic killing and to second order killing tensor fields. Together with the Hamiltonian constraint these four charges, when expressed in velocity phase space, provide an algebraic system for the two positions and the corresponding velocities; its solution is enough to reveal the entire solution space of the problem.
Except the well known solutions \eqref{solut1} and  \eqref{solut2}, the solution \eqref{euclsol1} is,to the best of our knowledge, new. When $\kappa_1<0$, it exhibits the interesting phenomenon of signature change as $b(t)$ spans the interval $(-\infty, \infty)$.

At the quantum level we have followed the canonical  quantization method. In order to determine a unique (up to constants) wave function, we have used-except the Wheeler-DeWitt constraint equation- the operators corresponding to Abelian subalgebras formed by the two pairs consisting of the before-said constraint and each of the two operators corresponding to the quadratic symmetry generators.

The linear non-local charge has been used as a guideline for selecting a natural physical time in the configuration space and thus define a conditional probability
with respect to which the classically singular configurations are assigned zero weight.

Last, but not least, we would like to discuss the existence of the notion of homothetic time adopted in this work. At first glance, one might think that this is particular to the situation depicted in the present work, i.e. the fact that we have a two dimensional configuration space; as a consequence every one- form is necessarily closed and therefore the same is true for the one-form corresponding to the homothetic vector. In fact this is not the case: the homothetic vector exists(as the only linear symmetry generator) even for the totally anisotropic case with $\gamma_{\ab}=diag(a^2(t),b^2(t),c^2(t))$; what is more important the corresponding one form is also closed. Thus, the described procedure of defining the conditional probability  adopted in this work will also be valid in this case as well. The same reasoning holds true even for a full scale factor matrix $\gamma_{\ab}$: the corresponding Langrangian density is $L=n \left( \frac{1} {R} L^{\abcd} \mathcal{K}_{\ab} \mathcal{K}_{\cd}-1\right)$,
where $\mathcal{K}_{\ab} \equiv \sqrt{\gamma} R K_{\ab}$, $K_{\ab}$ is the usual extrinsic curvature of the slice $t=$constant, $R$ is its Ricci scalar, $L^{\abcd}=\gamma^{\alpha\gamma}\gamma^{\beta\delta}+\gamma^{\alpha\delta}\gamma^{\beta\gamma}-2\gamma^{\ab}\gamma^{\gamma\delta}$, and we have redefined the usual lapse as $N=\frac{n}{\sqrt{\gamma}R}$ so that the vector $\gamma_{\ab}\frac{\partial}{\partial \gamma_{\ab}}$ is revealed as a homothecy generator for the scaled mini-supermetric $\mathcal{G}^{\abcd} \equiv \frac{\gamma R}{4} L^{\abcd}$. One can be satisfied that the corresponding one form is again closed. This may well be the starting point of future work.

\acknowledgments{\begin{figure}[h!]
\centering
  \begin{subfigure}[h]{0.265\linewidth}
    \includegraphics[width=\linewidth]{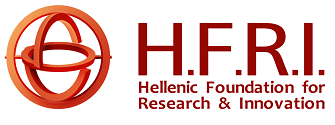}
  \end{subfigure}
  \begin{subfigure}[h]{0.2\linewidth}
    \includegraphics[width=\linewidth]{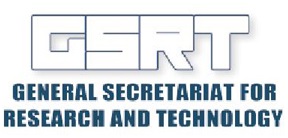}
  \end{subfigure}
\end{figure}
The research work was supported by the Hellenic Foundation for Research and Innovation (HFRI) and the General Secretariat for Research and Technology (GSRT), under the HFRI PhD Fellowship grant (GA.no.74136/2017). }

\end{document}